\documentclass[usenatbib]{mn2e}

\pdfoutput=1
\usepackage{amssymb,amsmath}
\usepackage{graphicx}
\usepackage{multirow}
\usepackage{natbib}
\usepackage{tabularx}

\def\cite{\citep}

\usepackage{url}
\usepackage{bm}
\footnotesize
\newdimen\minuswidth    
\setbox0=\hbox{$-$}
\minuswidth=\wd0
\catcode`@=\active
\def@{\kern\minuswidth}
\newdimen\digitwidth    
\setbox0=\hbox{\rm0}
\digitwidth=\wd0
\catcode`!=\active
\def!{\kern\digitwidth}
\normalsize

\usepackage[usenames,dvipsnames]{color}

\footnotesize
\newdimen\digitwidth    
\setbox0=\hbox{\rm0}
\digitwidth=\wd0
\catcode`!=\active
\def!{\kern\digitwidth}
\normalsize
\title[DM variations in 168 pulsars]{Dispersion measure variations in a sample of 168 pulsars}
\makeatletter
\author[E.~Petroff et al.]
{E.~Petroff$^{1,2}$\thanks{Email: epetroff@astro.swin.edu.au},
M.~J.~Keith$^2$,
S.~Johnston$^2$,
W.~van Straten$^1$,
R.~M.~Shannon$^2$
\\
$^1$ Centre for Astrophysics and Supercomputing, Swinburne University of Technology, P.O. Box 218, Hawthorn, VIC 3122, Australia \\
$^2$ CSIRO Astronomy \& Space Science, Australia Telescope National Facility, P.O. Box 76, Epping, NSW 1710, Australia\\
}
\makeatother

\date{}

\usepackage{float}

\begin{document}

\maketitle
\newcommand{\setthebls}{
}

\setthebls

\begin{abstract}
We analyse dispersion measure (DM) variations in six years of radio observations of more than 160 young pulsars, all gamma ray candidates for the \textit{Fermi} gamma ray telescope mostly located close to the Galactic plane. DMs were fit across 256~MHz of bandwidth for observations centred at 1.4~GHz and across three frequencies -- 0.7~GHz, 1.4~GHz, and 3.1~GHz -- where multifrequency observations were available. Changes in dispersion measure, dDM/d$t$, were calculated using a weighted linear fit across all epochs of available DMs. DM variations were detected at a 3$\sigma$ level in 11 pulsars, four of which were above 5$\sigma$: PSRs J0835$-$4510, J0908$-$4913, J1824$-$1945, and J1833$-$0827. We find that after 28 years of gradual decline, the DM of PSR J0835$-$4510 is now increasing. The magnitude of variations in three of the four (PSRs J0835$-$4510, J0908$-$4913, and J1833$-$0827) are above what models would predict for an ISM dominated by Kolmogorov turbulence. We attribute this excess as likely due to the pulsar's local environment - the supernova remnants near PSRs J0835$-$4510 and J1833$-$0827 and the pulsar wind nebula around PSR~J0908$-$4913. Upper limits were determined for all pulsars without detectable values of dDM/d$t$, most limits were found to lie above the levels of variations predicted by ISM theory. We find our results to be consistent with scattering estimates from the NE2001 model along these lines of sight.
\end{abstract}


\section{Introduction}

The emission from pulsars experiences a time delay as it passes through the interstellar medium (ISM) due to the dispersive effects of its plasma component. The group delay of this signal $t_g(\nu)$ depends on the observation frequency and the electron density along the line of sight as

\begin{equation}\label{eq:group}
t_g(\nu) = \frac{\left[\int_0^L n_e \mathrm{d}l \right]}{K \nu^2}
\end{equation}

\noindent where $K$ is the dispersion constant with a value $K \equiv 2.410 \times 10^{-4} \mathrm{\:MHz}^{-2} \mathrm{\:cm}^{-3} \mathrm{\:pc\: s}^{-1}$, $\nu$ is the observing frequency in GHz, $n_e$ is electron density and $L$ is the distance of the pulsar from the observer. The expression in brackets is refered to as the dispersion measure (DM) and describes the amount of ionised interstellar material between the observer and the pulsar \cite{lk+04}. 

Dispersive effects are proportional to $\nu^{-2}$ and can be determined experimentally by measuring delays in pulse arrival times for a pulsar across the bandwidth of an observation at a single frequency or fitting over a range of frequencies (\citet{kcs+13} and references therein). 

When a pulsar is first discovered, the spin period and DM are directly obtained as part of the search process.
In most cases, a pulsar's DM is treated as a constant but several long-term studies of DM have revealed temporal variations on timescales of months to years \cite{hhc85, pw91, yhc07, kcs+13}.

Variations in dispersion have been used to study turbulent structure in the free electron density of the ISM.
The spectral energy density scale of interstellar turbulence is thought to show power law statistics for interstellar material between large (10$^{18}$ m) and small (10$^6$ m) spatial scales such that

\begin{equation}\label{eq:powerLaw}
P_{3N}(q) \approx C_{N}^2 q^{-\beta}
\end{equation}

\noindent represents the power spectrum of the electron density $P_{3N}$ with a spectral index $\beta$ and a structure coefficient $C_{N}^2$. It is estimated in \citet{ars95} that this spectrum is consistent with a Kolmogorov power law described by $\beta = 11/3$. However, inhomogeneities in the form of highly anisotropic filaments are believed to exist in the ISM \cite{bmg+10} and may be responsible for so-called extreme scattering events \cite{fdj87,rbc87}. 

Turbulent interstellar material is probed on various scales by different types of observations, from rotation measure variations at the largest scales to weak diffractive interstellar scattering at the smallest. Fluctuations in pulsar DMs provide the capability to probe the ISM in the middle of this spatial range between $10^{11}$ and $10^{12}$ m, where other techniques are incapable of detecting variations. Thus, DM variation measurements bridge a crucial gap in the ISM turbulence spectrum.

Significant DM variations have previously been observed in studies of different classes of objects. Varying DM along the line of sight to the Vela pulsar was first noted in Hamilton et al. (1985); DM was observed to decrease over the length of their study. They attributed these changes to a dense, magnetised filament within the Vela supernova remnant (SNR) passing out of the line of sight over the 15 years of data \cite{hhc85}. Similarly, DM to the Crab pulsar has been observed to increase by 0.02 cm$^{-3}$ pc yr$^{-1}$ over 68 epochs between 1982 and 1988, attributed to variations within the turbulent environment of the local Crab SNR \cite{lps88}.
A study of seven pulsars over two years in \citet{pw91} detected variations caused by interstellar turbulence with a spectral index of $\beta = 11/3$, and DM variations have also been observed in high-precision observations of millisecond pulsars (e.g. \citealp{yhc07}).
These variations are generally consistent with levels of turbulence in the ionised ISM, though there is growing evidence that the exponent of the power-law noise process is steeper than expected from Kolmogorov turbulence for some lines of sight \cite{kcs+13}.

Changes in DM over time provide a direct method of probing turbulence in the interstellar medium \cite{r77}. These changes relate to the DM structure function $D_{\rm DM}$ as

\begin{equation}\label{eq:DDMtheory}
\left\lvert\frac{\mathrm{dDM}}{\mathrm{d}t} \right\rvert = \frac{(D_{\mathrm{DM}})^{1/2}}{\tau}
\end{equation}

\noindent where $\lvert$dDM/d$t \rvert$ is the absolute rate of change of the DM over time in cm$^{-3}$ pc yr$^{-1}$ and $\tau$ is the span of the observations in years \cite{bhv93}. The structure function, in turn, is related to the diffractive timescale, $\tau_d$, of the pulsar by \citet{yhc07}

\begin{equation}\label{eq:theory2}
D_{\mathrm{DM}} = \left(\frac{K \nu}{2 \pi}\right)^2 \left(\frac{\tau(s)}{\tau_d} \right)^{\alpha}  .
\end{equation}

\noindent $K$ has the same value as in Equation~\ref{eq:group}, $\tau(s)$ is the time span of the observations in seconds, and $\alpha = \beta - 2$, where $\beta = 11/3$ is taken to be the power-law exponent of a Kolmogorov spectrum \cite{ars95}. Here $\tau_d$ is the diffractive timescale in seconds at the observing frequency $\nu$. 

In this paper we examine the DMs of more than 160 young, highly energtic pulsars monitored regularly over six years at the Parkes radio telescope as part of a radio counterpart study to one conducted with the \textit{Fermi} gamma ray telescope \cite{sgc+08}. 
The pulsars in our sample are distributed at a range of distances within the Galactic plane with DMs between 2 cm$^{-3}$ pc and almost 1000 cm$^{-3}$ pc. The majority are at low Galactic latitudes and probe a diverse range of sight lines through the Galactic ISM.
Our sample is additionally promising for DM variation studies as young pulsars are also more likely to be associated with supernova remnants remaining from their birth, providing the possibility of yet more ionised, turbulent local structure.

Many previous studies mentioned here focus on millisecond pulsars, most with DM $<$ 100 cm$^{-3}$ pc.
Canonical pulsars are detected up to much higher DMs and provide a complementary sample to the MSPs, allowing us to study turbulence over larger lines of sight.

In Section \ref{sec:observations} we describe our observations; in Section \ref{sec:methods} we describe our data analysis procedures and our statistic for measuring DM variations in our pulsars. In Section \ref{sec:results} we outline our findings, with special attention paid to four pulsars of interest: PSRs J0835$-$4510, J0908$-$4913, J1824$-$1945, and J1833$-$0827, and we set upper limits for all others in Section \ref{sec:upperLimits}; we conclude in Section \ref{sec:conclusions}.

\section{Observations}\label{sec:observations}
Since early 2007, regular observations of a large sample of pulsars have been carried out with the 64-metre Parkes
radio telescope in support of the \textit{Fermi} gamma-ray mission. A total of 156 pulsars were
drawn from the list of highly energetic pulsars in Smith et al. (2008) supplemented by a small number of other
interesting southern sources. The initial description of the observational setup and early timing results from 
the Parkes dataset are described in \citet{wjm+10}.

For this paper we used data taken using the Parkes telescope between February 2007 and October 2012. Observations were carried out on an approximately monthly basis with all 168 pulsars observed over a 24 hour period at a centre frequency near 1.4~GHz. At 6 month intervals, additional observations were obtained simultaneously at 3.1 and 0.7~GHz on the following day. During a single observation each pulsar was observed long enough to obtain a signal to noise ratio greater than 5, typically only a few minutes. 

Observations at 1.4~GHz with 256~MHz of bandwidth were taken using the centre beam of the Parkes multibeam receiver \cite{swb96}. Observations at 3.1 and 0.7~GHz were done simultaneously using the 10/50 cm receiver installed at Parkes \cite{gzf+05} and had 1024~MHz and 64~MHz of bandwidth, respectively \cite{wjm+10}.

The voltage signals from the orthogonal, linearly-polarised receptors were digitised and converted to a filterbank consisting of 1024 frequency channels then folded using 1024 phase bins across the pulse period of the pulsar. Incoming data were folded at the period of the pulsar in 30-second sub-integrations and then written to disk. Regular calibration was performed by sending an artificial calibration signal into the feed at a 45$^{\circ}$ angle from both linear probes to determine both their relative gain and the phase offset.
The final calibrated data were saved to disk and used to create an average pulse profile over the full span of the observation. 

\section{Analysis}\label{sec:methods}

Initial data reduction was performed after each observation using the {\sc psrchive} data analysis package \cite{hvm+04}. An integrated pulse profile for each pulsar observation was produced after excising time and frequency channels with significant radio frequency interference. The remaining channels were calibrated to produce a final pulse profile at each observing frequency which was compared with a standard profile for the pulsar, created by summing profiles from all previous observations \cite{wj+08}. 

The time of arrival (TOA) for the integrated profile was converted to the solar system barycentre using the DE405 model \cite{s98} and compared with that of the timing model prediction using the {\sc tempo2} software package \cite{hem+06}. The time difference between the observed TOA and the timing model gave a residual for each observation. Once several epochs of residuals were available, it became possible to remove common parameters such as effects from the pulsar spin frequency $\nu$ and spin frequency derivative $\dot{\nu}$ through fits within {\sc tempo2}. These fits removed linear and quadratic terms from the residuals, respectively. At this stage it was possible to fit for the DM contribution in the refined residuals. 

Previous studies of DM from pulsar timing residuals have identified problems of obtaining reliable values and accurately correcting for DM effects in the timing solution. You et al. (2007) performed a linear fit by directly comparing the arrival times at two different frequencies. Since all epochs in their sample consisted of observations at 3 distinct centre frequencies they determined the best two with which to perform calculations on a case-by-case basis. This method was updated in \citet{kcs+13} by using all three observing frequencies simultaneously to determine the DM for each epoch. Keith et al. also incorporated a smoothing function into the {\sc tempo2} fitting algorithm that was previously applied \textit{after} fitting. 

We created two DM datasets using the algorithm described in \citet{kcs+13}. Each pulsar has been observed at approximately 80 epochs at 1.4~GHz but at only 10 epochs at 0.7 and 3.1~GHz.
Thus two separate time series were created, one for DM measured across the 20 cm band only, and one measured across all three frequencies, where available.
In some cases the number of useful multifrequency DM measurements is smaller than the number of epochs at which observations were taken because scattering effects preclude detection at 0.7~GHz.

To calculate DM variations over the span of the dataset, we performed a weighted least-squares fit to measure the slope in DM over time. The result of this fitting procedure was a best-fit slope, dDM/d$t$, taken to be the variation in DM. Separate fits were done for 20 cm and multifrequency DM measurements resulting in two separate dDM/d$t$ values for each pulsar. Multifrequency fits of dDM/d$t$ for pulsars with only one or two usable multifrequency epochs were not considered as the linear fit had no associated error.

We note that the weighted linear fit is not a perfect tool for the study of small-scale variations, and we may expect variations on timescales shorter than the many-year span of our dataset that do not fit this trend. Overarching linear changes in DM would be expected to arise in cases where a single large structure moves across the line of sight over the span of observations. Additionally, the steep power law of Equation~\ref{eq:powerLaw} implies that the most power will be at the largest timescale, thus the largest DM variations. A linear fit may also encompass spatial density gradients but provides a good first order detection statistic.

\section{Results - Detections}\label{sec:results}

\begin{table}\caption{Pulsars with DM variations over 6 years above 3$\sigma$ levels. Pulsars with detections above 5$\sigma$ levels are listed in bold. Slopes for the multifrequency (mf) and 20 cm observations with errors in the last digit are in units of cm$^{-3}$ pc yr$^{-1}$. \label{tab:significant}}
\begin{centering}
\begin{tabular}{l c c c}
\hline
Name & DM & dDM/d$t_{mf}$ & dDM/d$t_{20cm}$ \\
\hline
\textbf{J0835$-$4510} 	& 67.9 	& 0.005(1) 	& 0.0081(9) \\
\textbf{J0908$-$4913} 	& 180.4 	& $-$0.038(4) 	& $-$0.030(1) \\
\textbf{J1824$-$1945} 	& 224.4 	& $-$0.011(2) 	& $-$0.022(1) \\
\textbf{J1833$-$0827} 	& 411 	& $-$0.13(2) 	& $-$0.18(1) \\
J0834$-$4159		& 240.4 	& $-$0.020(5)	& $-$0.27(3) \\
J1112$-$6103 	& 599.0 	& $-$0.11(4) 	& $-$0.52(5) \\
J1702$-$4128 	& 366.7 	& 0.12(3) 	& 0.4(1) \\
J1721$-$3532  	& 497.01 	& $-$0.047(9) 	& $-$0.19(6) \\
J1745$-$3040 	& 88.112 	& $-$0.018(2) 	& $-$0.016(3) \\
J1809$-$1917 	& 196.9 	& 0.07(2) 	& 0.17(4) \\
J1826$-$1334 	& 230.8 	& 0.14(4) 	& 0.13(3) \\
\hline
\end{tabular}
\end{centering}
\end{table}
DM variations were detected in eleven pulsars from our sample over the 6 years of observation. All are embedded in the Galactic plane with dispersion measures ranging from 67.9 to 599 cm$^{-3}$ pc. Only four pulsars in our sample, PSRs J0835$-$4510, J0908$-$4913, J1824$-$1945, and J1833$-$0827, had variations deemed highly significant. These pulsars wer identified based on the fact that they all had values of dDM/d$t$ with agreement in sign between 20 cm and multifrequency fits with an error in each fit $\leq 35\%$. Variations were labeled as highly significant detections if errors were $\leq 20\%$ in both fits. Fits for all other pulsars in our sample failed to meet one or more of these criteria and the weighted linear fit was used to produce an upper limit on detectable variations. 

Marginal detections, pulsars with detections between 3$\sigma$ and 5$\sigma$, are largely consistent with variations predicted from an ISM dominated by Kolmogorov turbulence.

Table~\ref{tab:significant} lists the pulsars with significant measurements
of DM variations.Highly significant pulsars will be discussed individually below.

\subsection{PSR J0835$-$4510}
PSR J0835$-$4510 (B0833$-$45), located in the Vela supernova remnant, lies within the Gum Nebula \cite{lvm68}. It is one of the brightest pulsars in the sky, with a flux at 1400~MHz of 1100 mJy \cite{bf74} and characteristic age $\tau_c$ of 11 kyr. It is located at a distance from the sun of approximately 300 pc \cite{dlr+03} at Galactic longitude and latitude $(\ell, b) = (263.5^{\circ}, -2.79^{\circ})$. 
The pulsar has a very large DM for its distance and using $C_{N}^2$ as a measure of turbulence \cite{c86},
its value is the highest measured of any pulsar \cite{jnk98}.

\begin{figure}
\includegraphics[width=6cm,angle=-90]{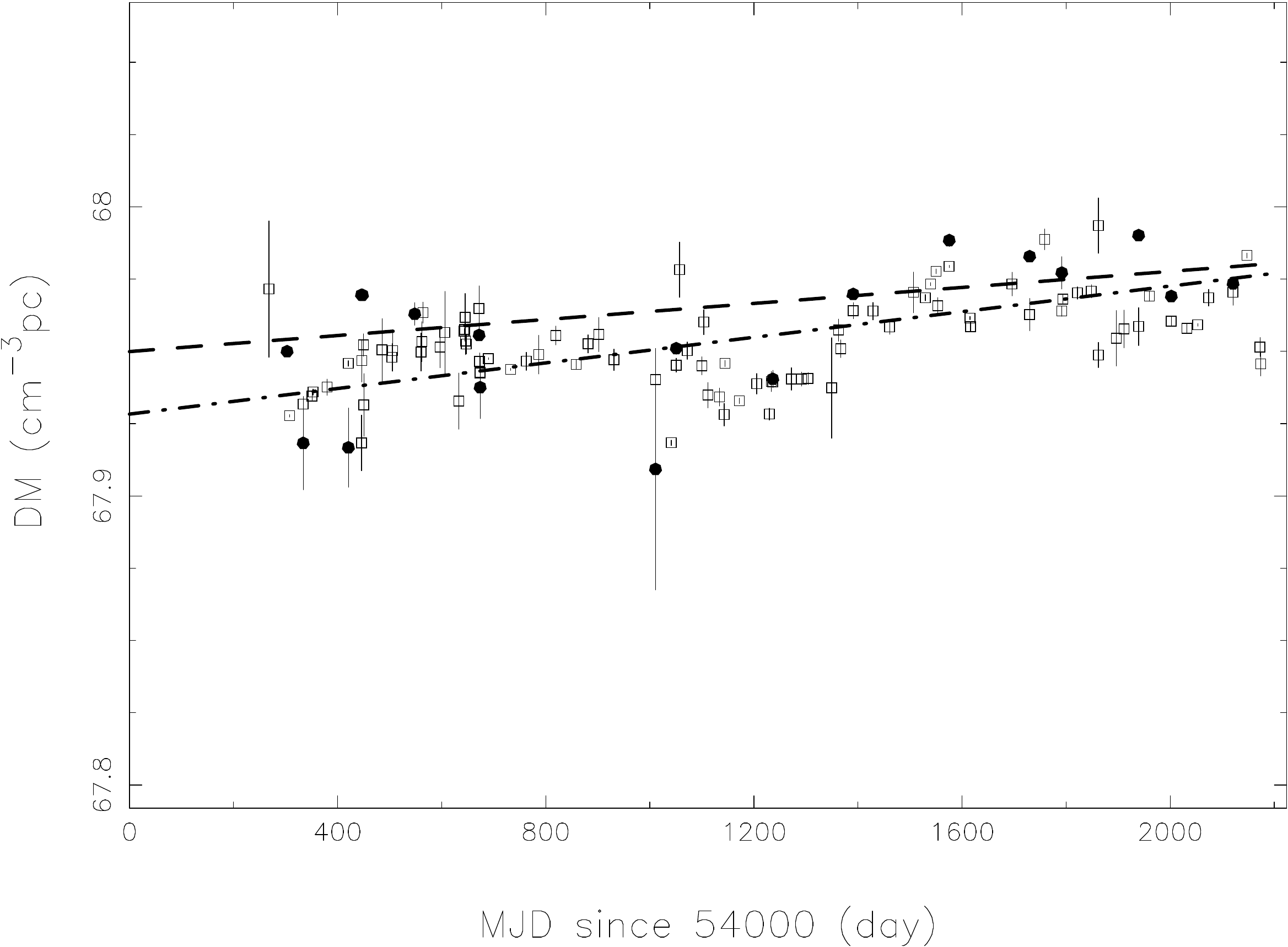}
\centering
\caption{The DM of PSR J0835$-$4510 over 2000 days for 18 epochs of multifrequency measurements (circles) and 97 epochs of measurements across the band centred at 20 cm (squares) with weighted linear fits represented by the dashed and dot-dashed lines, respectively. The general trend is of increasing DM over the dataset with a notable reversal between MJDs 54600 and 55200. \label{fig:Vela}}
\end{figure}

The scintillation velocity at high observing frequencies has been measured by
\citet{jnk98}. Corresponding to a value of $\tau_d\sim$10~s at 1.4~GHz. We therefore expect
DM variations of order 0.01 cm$^{-3}$ pc yr$^{-1}$ based on Equation~\ref{eq:DDMtheory}.
Previous studies of changing DM along the line of sight to Vela \cite{hhc85} showed DM to be decreasing over 5 epochs between the years 1970 and 1985 at a rate of 0.040 cm$^{-3}$ pc yr$^{-1}$, a factor of 4 greater than expected from turbulence. \citet{hhc85} interpreted their measured decrease in DM as the movement of a magnetised filament out of the line of sight within the SNR, meaning that changes in DM were due to the local environment of the pulsar rather than the turbulence in the greater ISM. 

As shown in Figure \ref{fig:Vela}, single frequency DM measurements over the entire six years of our dataset show DM \textit{increasing} at a rate of 0.0081(9) cm$^{-3}$ pc yr$^{-1}$. Similarly, dDM/d$t$ = 0.005(1) cm$^{-3}$ pc yr$^{-1}$ using multifrequency fits.
These variations are significantly smaller than those measured by Hamilton et al. (1985) but are more or less consistent with expected values given the measured scintillation parameters.

\begin{figure}
\includegraphics[width=6cm,angle=-90]{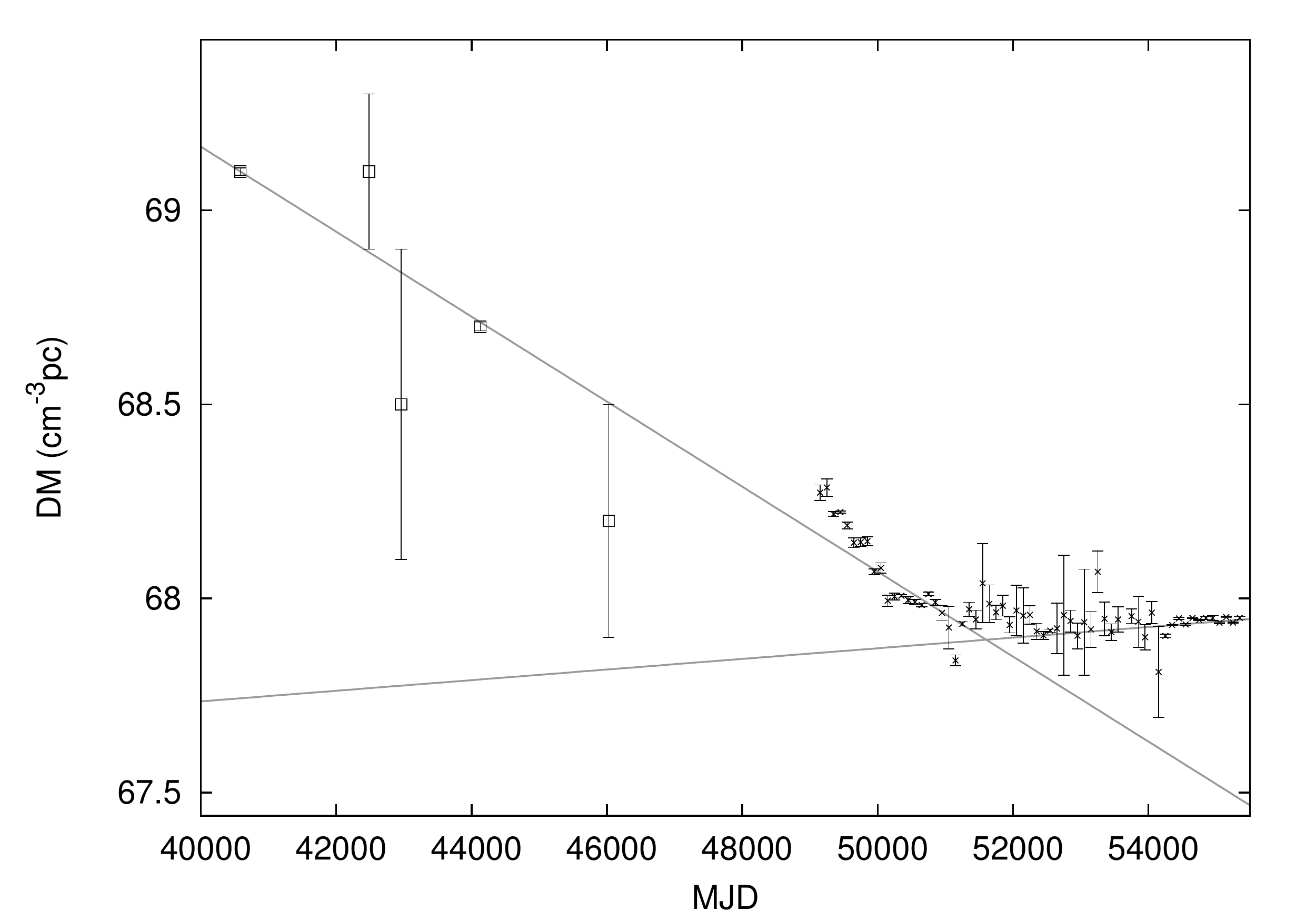}
\centering
\caption{DM measurements of PSR J0835$-$4510 since 1969 (MJD 40500). Square markers indicate single values taken from Hamilton et al. (1985). Other markers are weighted averages of the DM measured in 100 day bins of Parkes observations. Values before MJD 54200 are taken from archival Parkes data recorded using an analogue filterbank system. The lines indicate extrapolation of the trends from the Hamilton et al. dataset and from this work. \label{fig:vela_archival}}
\end{figure}

The dramatic change in dDM/d$t$ is highlighted in Figure ~\ref{fig:vela_archival} where we show the Hamilton et al. data over plotted with data from this study and analogue filterbank system archives at Parkes.
In order to reduce the noise due to measurement uncertainty we plot weighted averages in 100-day bins.
We caution that there may be a systematic offset between the DMs derived in the Hamilton et al. data and those derived from our work due to known intrinsic frequency evolution of the Vela profile \cite{kjl+11}.
We interpret the dramatic change in dDM/d$t$ to be evidence that the filament responsible for change seen by Hamilton et al. moved completely out of our line of sight some time near MJD 50000. We believe that the currently observable DM variations can be explained solely by the turbulent ISM.

Although the overall trend in dDM/d$t$ in the recent data is positive, variations are visible on shorter timescales within the span of our observations, most notably where DM appears to decrease between MJD 54600 and 55200. 
This is consistent with turbulence in the ISM. Contributions from the surrounding Gum Nebula are most likely minimal, however, as there are no detectable variations in other pulsars from the Gum in our sample -- PSRs J0738$-$4042, J0742$-$2822, J0745$-$5351, and J0905$-$5127.

\subsection{PSR J0908$-$4913}
The pulsar PSR J0908$-$4913 (B0906$-$49) has a spin period of 106~ms, and
a DM derived distance of 6.7~kpc placing it well within the Galactic plane, at $(\ell, b) = (270.27^{\circ}, -1.02^{\circ})$ \cite{dmd88}.
\citet{gsf98} discovered a bow-shock in the immediate surroundings of the pulsar suggesting that it is travelling through the turbulent medium of an associated pulsar wind nebula (PWN) with velocity ~60 km s$^{-1}$ at a position angle of 315$^{\circ}$ (north through east).
Their derived velocity is consistent with the scintillation measurements of Johnston et al. (1998), who measured the diffractive timescale to be 4770~s at 1.5~GHz.
This yields expected DM variations of $6.0 \times 10^{-5}$ cm$^{-3}$ pc yr$^{-1}$.

No long term studies of DM along the line of sight to PSR J0908$-$4913 exist in the literature, however previously published DM estimates for this pulsar do exist as far back as its discovery in 1988, when its DM was measured to be 192 $\pm$ 12 cm$^{-3}$ pc \cite{dmd88}. At the beginning of the \textit{Fermi} dataset in 2007 DM was measured to be approximately 180.42 cm$^{-3}$ pc, but had dropped by 0.12 cm$^{-3}$ pc by the final epoch from 2012.
Unfortunately the large uncertainty in the 1988 data and lack of data in the intervening years prevent us from drawing any conclusions as to the long term evolution of the DM.

\begin{figure}
\centering
\includegraphics[width=6cm,angle=-90]{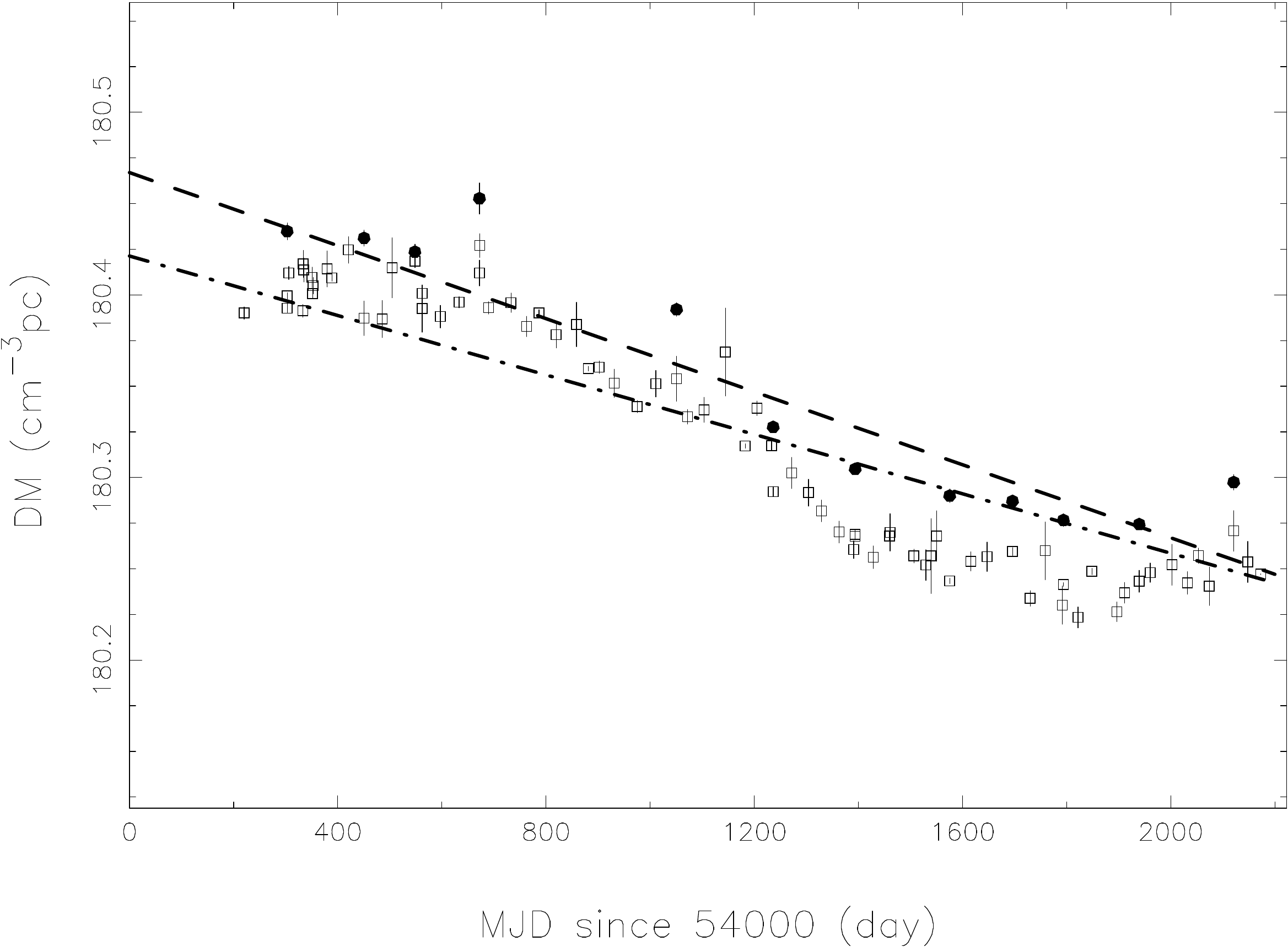}
\caption{The DM of PSR J0908$-$4913 over 12 multifrequency epochs (circles) and 77 epochs measured across only the 20 cm band (squares) with weighted linear fits represented by the dashed and dot-dashed lines, respectively. \label{fig:J0908}}
\end{figure}

The variation in DM along the line of sight to PSR J0908$-$4913 is one of the largest measured of any pulsar in our sample, with a best-fit linear slope of dDM/d$t = -0.030(1)$ cm$^{-3}$ pc yr$^{-1}$ for the 20 cm DM measurements and dDM/d$t = -0.038(4)$ cm$^{-3}$ pc yr$^{-1}$ over 12 multifrequency epochs shown in Figure~\ref{fig:J0908}.
Similar to PSR J0835$-$4510, fluctuations on shorter timescales are present in the 20 cm DM data, and are visible even in the more sparsely sampled multifrequency dataset.

The DM variations along the line of sight are more than two orders of magnitude greater than expected from models.
The high variations, then, seem likely to be caused by the immediate
surroundings of the pulsar. PSR~J0908$-$4913 has an unusual PWN and appears to be moving slowly through a highly dense ($n > 2$~cm$^{-3}$), and likely turbulent, medium \cite{gsf98}. Thus for a region on the order of a parsec in size, the PWN contribution to total DM would be small but the nebula's highly turbulent nature would be capable of much larger fractional contributions to dDM/d$t$ as the pulsar moved through it. DM variations can then be attributed to a small percentage variation from within the local PWN.

\subsection{PSR J1824$-$1945}

PSR J1824$-$1945 (B1821$-$19) is a young pulsar with a period of 189~ms and a characteristic age $\tau_{c} = 5.7 \times 10^5$ yr \cite{mlt78}. It lies in the Galactic plane close to the centre of the Galaxy with Galactic coordinates $\ell = 12.28^{\circ}$ and $b=-3.11^{\circ}$ at a DM-derived distance of approximately 5.2 kpc \cite{cl02}.

Previous studies of PSR~J1824$-$1945 were unable to make significant measurements of pulsar velocity or DM variations \cite{zhw+05,hlk+04}, thus neither significant velocity nor diffractive timescale measurements exist for this pulsar in the literature. Assuming a velocity of 300 km~s$^{-1}$ we expect a diffractive timescale of order $\tau_d$ = 6~s, which in turn corresponds to DM variations of magnitude 0.02 cm$^{-3}$ pc yr$^{-1}$ using the \citet{cl02} model.

\begin{figure}
\centering
\includegraphics[width=6cm,angle=-90]{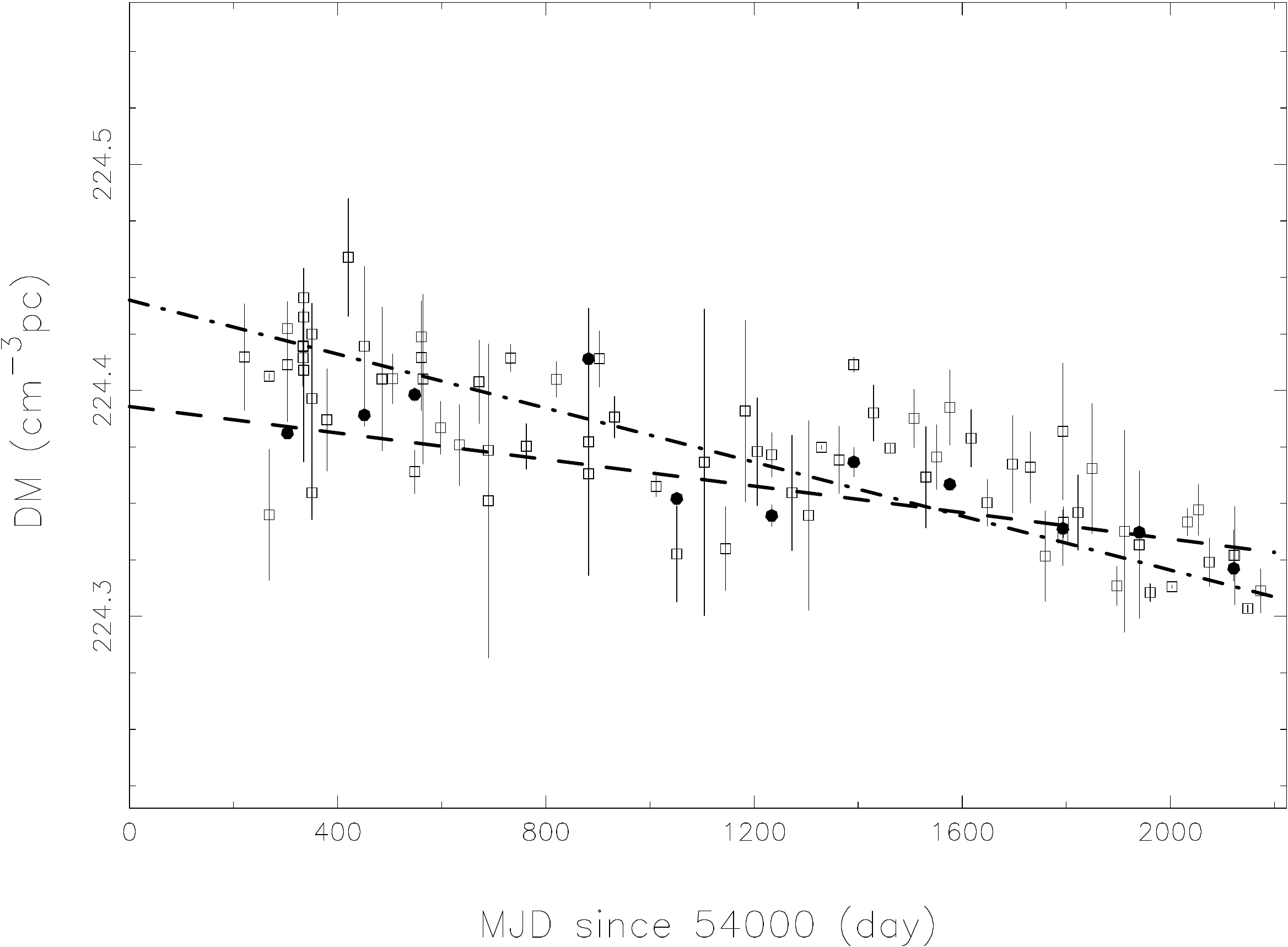}
\caption{DM measurements for PSR J1824$-$1945 over 11 epochs of multifrequency data (circles) and 75 epochs measured across the 20 cm band alone (squares) with weighted linear fits represented by the dashed and dot-dashed lines, respectively. \label{fig:J1824}}
\end{figure}

The general trend in our data is a decrease in DM over the six years of observations seen in Figure~\ref{fig:J1824}, although the best-fit rate of this variation differs between the 20 cm and the multifrequency data with dDM/d$t_{20cm}= -0.022(1)$ cm$^{-3}$ pc yr$^{-1}$ and dDM/d$t_{mf} = -0.011(2)$ cm$^{-3}$ pc yr$^{-1}$, respectively. These gradients differ by a factor of 2, with lower error in the fit at 20 cm. The good fit to the data may arise because the trend over 2000 days is not a strictly linear one; additional fluctuations are visible on shorter timescales possibly due to turbulent structure on small scales passing through the line of sight. 

In the case of PSR J1824$-$1945 the DM variations we observe are consistent with values predicted for a turbulent ISM in this direction for our assumed velocity and DM-determined distance. This pulsar is not well-studied like PSRs J0835$-$4510 and J0908$-$4913, but there is no evidence for the presence of an associated SNR or PWN in the local neighbourhood. Therefore we attribute the variations in the DM towards this pulsar to the turbulent ISM alone.

\subsection{PSR J1833$-$0827}

PSR J1833$-$0827 (B0830$-$08) has a period of 85 ms and an approximate characteristic age of ~150 kyr \cite{cl86}. At the time of its discovery it was identified as an isolated pulsar close to the supernova remnant W41. It has been argued that the pulsar's high velocity away from SNR W41 indicates a past connection and that PSR J1833$-$0827 originated within the shell-like SNR \cite{hll+05}. More recently, X-ray studies of the region using the XMM-Newton satellite discovered an X-ray pulsar wind nebula surrounding this pulsar in the form of a bow shock nebula \cite{eit+11}.

\begin{figure}
\centering
\includegraphics[width=6cm,angle=-90]{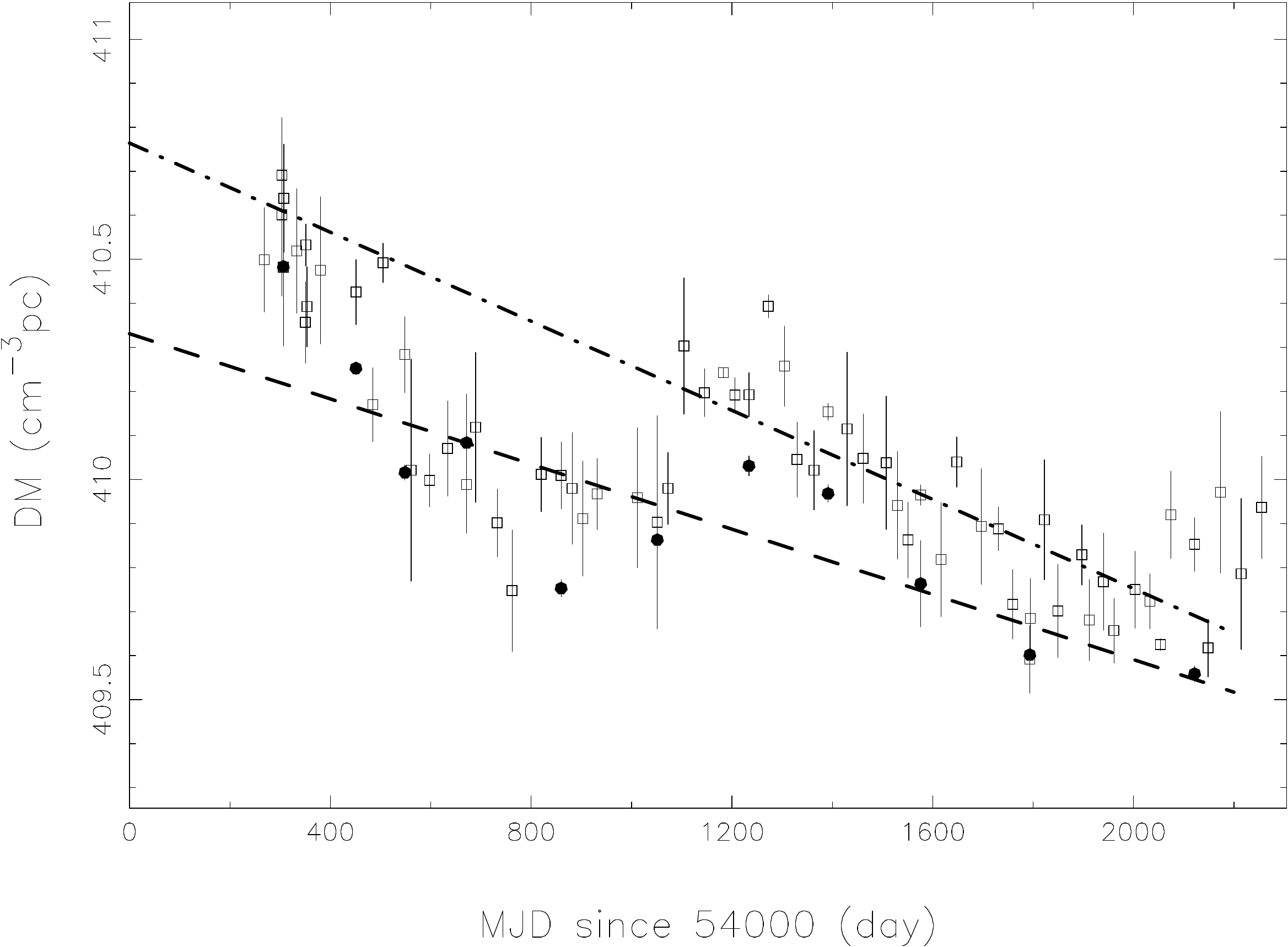}
\caption{The DM of PSR J1833-0827 over the 2000 days of observations as measured over 11 epochs of multifrequency data (circles) and 67 epochs measured across the 20 cm band alone (squares) with weighted linear fits represented by the dashed and dot-dashed lines, respectively. The general trend is that of decreasing DM with a  sign reversal between MJD 54900 and 55300. \label{fig:J1833}}
\end{figure}

PSR J1833$-$0827 is located towards the Galactic centre ($\ell$, $b$) = ($23.39^{\circ}$, $0.06^{\circ}$) at a DM-derived distance of 5.6 kpc \cite{cl02}. This pulsar has been observed to move with a transverse velocity of approximately 740 km s$^{-1}$, more than twice the rms velocity of non-millisecond pulsars \cite{hll+05, nj95}. The high transverse velocity and distance correspond to a scattering timescale at 1400~MHz of 5 s and expected DM variations of 0.02 cm$^{-3}$ pc yr$^{-1}$.

From our observations we find a general trend of decreasing DM along the line of sight to J1833$-$0827 with dDM/d$t_{mf} = -0.13(2)$ and dDM/d$t_{20cm} = -0.18(1)$ in units of cm$^{-3}$ pc yr$^{-1}$, seen in Figure~\ref{fig:J1833}, for multifrequency and 20 cm DMs, respectively. These variations are larger than those of any other pulsar in our sample by an order of magnitude. However, these fits include a change in sign of dDM/d$t$ between MJD 54900 and MJD 55300 seen in both sets of DM measurements, most likely attributable to general turbulence; a piecewise fit to these data would yield an even larger value for dDM/d$t$ over the regions of decreasing DM. PSR J1833$-$0827 also has the largest DM of any pulsar in which variations were detected. 

These extreme DM variations may seem more reasonable in light of the recent discovery of the pulsar's X-ray PWN. The turbulent ISM along the line of sight to PSR J1833$-$0827 would be expected to contribute only about 10\% of the observed variations and it is possible that observed behaviour is due entirely to the energetic bow shock nebula through which the pulsar is moving, including the brief passage of a dense filament through the line of sight corresponding to the temporary increase in DM midway through the dataset.

\section{Results - Upper limits}\label{sec:upperLimits}

\begin{figure*}
\centering
\includegraphics[width=12cm,height=9cm]{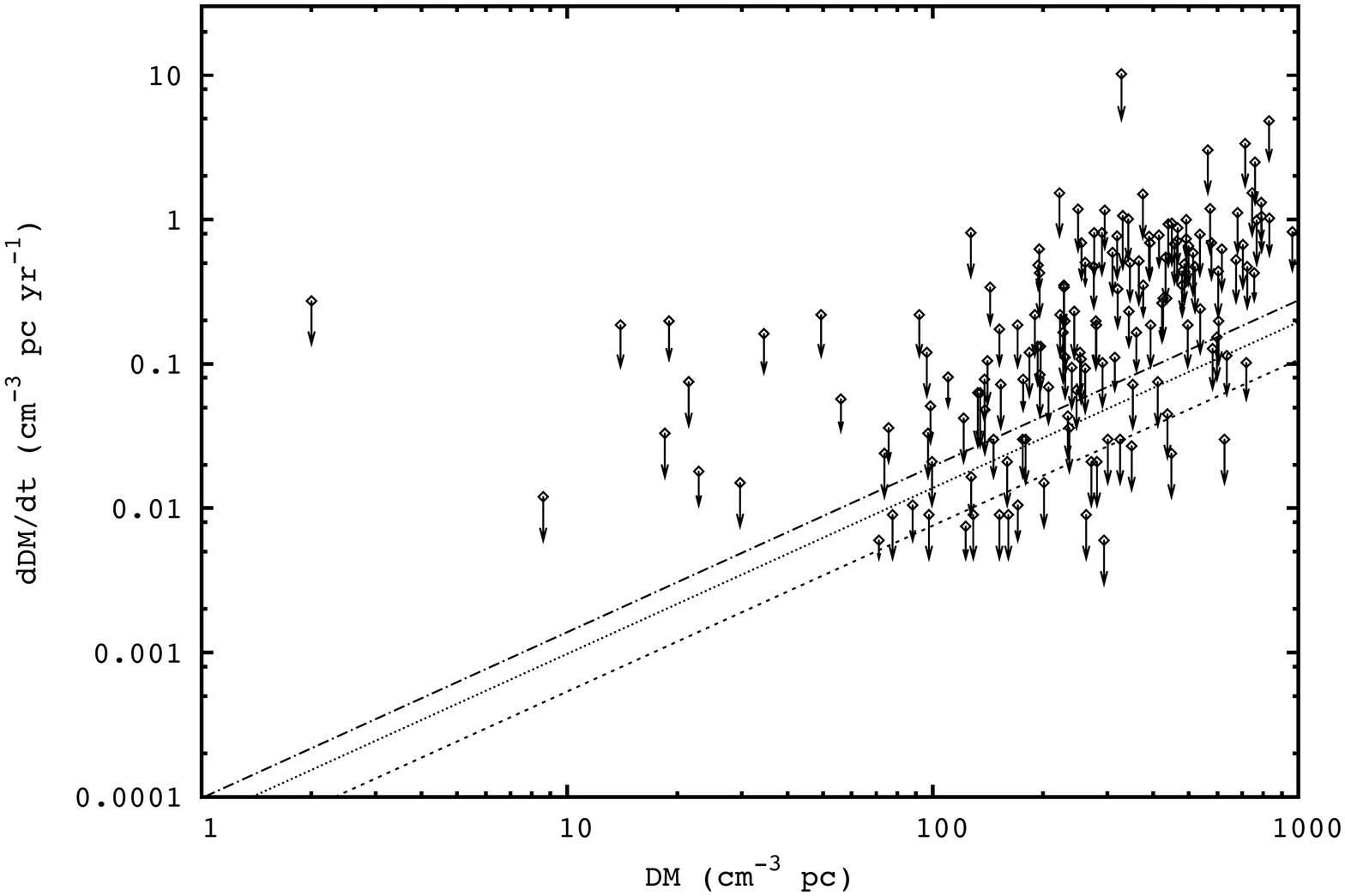}
\caption{Upper limits on dDM/d$t$ for pulsars in which no significant DM variations were detected. Overlayed lines display predicted variations detectable at a range of DMs for a pulsar at $\ell = 330^{\circ}$, $b = 0^{\circ}$ with a velocity of 164 km~s$^{-1}$ (dashed), 338 km~s$^{-1}$ (dotted), and 511 km~s$^{-1}$ (dot-dashed), the median and 1 sigma velocities from the pulsar velocity distribution in \citet{hll+05}. \label{fig:theory}}
\end{figure*}

Upper limits on dDM/d$t$ for each of the pulsars in the \textit{Fermi} project with no significant DM variations are listed in Table~\ref{tab:limits} along with the DM that best fits our data. All but 36 of our measured DM values have smaller uncertainties than previous best estimates.
In Figure~\ref{fig:theory} we plot our upper limits as a function of DM. We compared the 25 pulsars common to our observations and the \citet{hlk+04} study and found our upper limits were consistent with but less constraining than theirs.

 Although our limits on dDM/d$t$ may also contain contributions from spatial density gradients our data are not sufficiently sensitive to disentangle this contribution from that of variations due to turbulence. Because we are only setting upper limits on total variations, we do not attempt to explicitly differentiate between the two. Ideally we want to compare our upper limits with theoretical 
predictions for dDM/d$t$ based on Kolmogorov turbulence.
To do this we would need to measure the diffractive timescale $\tau_d$ and then apply
Equations~\ref{eq:DDMtheory} and~\ref{eq:theory2}. Unfortunately, the
high DMs of most of our sample preclude direct measurements of $\tau_d$
as, in the majority of cases, they will be significantly smaller than
our integration time.
For any given line of sight we can however, estimate $\tau_d$ using
the Cordes \& Lazio (2002) model of the Galaxy, which estimates scintillation along the line of sight consistent with a Kolmogorov spectrum of turbulence. For illustrative purposes in Figure~\ref{fig:theory} we choose a representative line of sight
at ($\ell$, $b$) = ($330^{\circ}$, $0^{\circ}$) and use
the Cordes \& Lazio (2002) model to step through DM and derive the 
scintillation bandwidth, $\nu_{d}$.
Conversion of $\nu_{d}$ to $\tau_{d}$ is obtained through

\begin{equation}\label{eq:td}
\tau_d = A_V \frac{\left(D \nu_d\right)^{1/2}}{\nu \:v},
\end{equation}

\noindent where $D$ is the distance to the pulsar in kpc, $\nu_d$ is the diffractive 
scintillation bandwidth in MHz at the observing frequency $\nu$ in GHz,
$v$ is the velocity of the pulsar in km~s$^{-1}$, and we adopt a value
of $A_V = 3.85 \times 10^4$ from previous studies \cite{nnd01}.
Figure~\ref{fig:theory} shows the curves obtained for three different
velocities, $v$= 164, 338, and 511 km~s$^{-1}$, the median and one sigma 2D velocities obtained using the distribution in \citet{hll+05}.

\begin{table*}\caption{All pulsars observed for the \textit{Fermi} project are listed alongside DM fits from our data with error in the last digit, and an upper limit on dDM/d$t$ in cm$^{-3}$ pc yr$^{-1}$. DM values with smaller uncertainties than previous best estimates are marked with an asterisk (*).}\label{tab:limits}
\begin{tabular}[H]{l c c | l c c}
\hline
Name & DM & dDM/d$t_{\rm lim}$ & PSR & DM & dDM/d$t_{\rm limit}$ \\
\hline
J0108$-$1431 	& 1(13) 	& 0.18 	& J1513$-$5908 	& 255.0(3) 	& 0.46 \\
J0401$-$7608* 	& 21.7(1) 	& 0.05 	& J1514$-$5925* 	& 194.0(4) 	& 0.3 \\
J0536$-$7543*	 & 18.58(2) 	& 0.2 	& J1515$-$5720* 	& 480.6(4) 	& 0.2 \\
J0543+2329	 & 77.556(5) 	& 0.006 	& J1524$-$5625* 	& 152.2(1)	 & 0.1 \\
J0614+2229* 	& 96.91(4) 	& 0.02 	&  J1524$-$5706* 	& 832(1) 	& 0.7 \\
J0627+0705*	 & 138.25(7) 	& 0.05 	&  J1530$-$5327* 	& 49.6(1) 	& 0.1\\
J0630$-$2834	 & 34.4(1)	 & 0.1 	&  J1531$-$5610* 	& 110.41(3) 	& 0.05 \\
J0659+1414*	 & 13.94(9)	 & 0.1 	& J1538$-$5551* 	& 604.6(1) 	& 0.3  \\
J0729$-$1448*	 & 91.7(2)	 & 0.1 	&  J1539$-$5626* 	& 175.85(3) 	& 0.02 \\
J0738$-$4042* 	& 160.896(3)	 & 0.006 	& J1541$-$5535* 	& 426.1(1) 	& 0.2  \\
J0742$-$2822* 	& 73.728(1) 	& 0.02	 & J1543$-$5459* 	& 345.9(3) 	& 0.3 \\
J0745$-$5353* 	& 121.38(2) 	& 0.03 	& J1548$-$5607* 	& 314.66(7) 	& 0.07 \\
J0821$-$3824 	& 195.8(4) 	& 0.3 	&  J1549$-$4848 	& 55.94(4) 	& 0.04\\
J0855$-$4644*	 & 236.4(1)		& 0.02	&  J1551$-$5310* 	& 491.6(7) 	& 0.7  \\
J0857$-$4424	 & 183.4(1) 	& 0.08 	& J1600$-$5044*	 & 262.791(4) 	& 0.006  \\
J0901$-$4624* 	& 199.3(2)	 & 0.2	&  J1600$-$5751 	& 176.4(1) 	& 0.05\\
J0905$-$5127 	& 196.21(4) 	& 0.06 	&   J1601$-$5335*	 & 195.2(6) 	& 0.4 \\
J0940$-$5428* 	& 134.55(3)	 & 0.04 	& J1602$-$5100* 	& 170.79(1) 	& 0.007  \\
J0954$-$5430* 	& 201.57(5) 	& 0.01	 &  J1611$-$5209* 	& 127.345(9) 	& 0.01 \\
J1003$-$4747* 	& 98.49(8) 	& 0.0001 	&  J1614$-$5048* 	& 582.4(1) 	& 0.09 \\
J1015$-$5719* 	& 278.1(2) 	& 0.1  	& J1632$-$4757* 	& 574.2(5)	 & 0.8 \\
J1016$-$5819* 	& 252.16(7) 	& 0.08	& J1632$-$4818 	& 758(5) 	& 0.3 \\
J1016$-$5857* 	& 394.48(9)	 & 0.1 	& J1637$-$4553 	& 194.7(1) 	& 0.09\\
J1019$-$5749* 	& 1040(1) 	& 0.8 	& J1637$-$4642* 	& 419.1(3) 	& 0.5 \\
J1020$-$6026*	 & 441.5(4)	 & 0.6 	& J1638$-$4417* 	& 436.4(3) 	& 0.2 \\
J1028$-$5820 	& 96.506(2) 	& 0.006 	& J1638$-$4608* 	& 423.1(1) 	& 0.2 \\
J1043$-$6116*	& 448.91(2) 	& 0.02  	& J1640$-$4715* 	& 586.32(6) 	& 0.5 \\
J1048$-$5832* 	& 128.679(4) 	& 0.006	& J1643$-$4505* 	& 478.6(6) 	& 0.3 \\
J1052$-$5954* 	& 491.9(6) 	& 0.5 	& J1646$-$4346 	& 499.2(3)	 & 0.4 \\
J1055$-$6032* 	& 636.5(1) 	& 0.07 	& J1648$-$4611* 	& 392.3(3) 	& 0.4 \\
J1057$-$5226* 	& 29.69(1)	 & 0.01	& J1649$-$4653* 	& 331(1) 	& 0.7  \\
J1105$-$6107* 	& 271.24(1) 	& 0.01 	&J1650$-$4502* 	& 320.2(3) 	& 0.2 \\
J1115$-$6052*	 & 226.92(5) 	& 0.1 	& J1650$-$4921* 	& 229.3(3) 	& 0.2 \\
J1119$-$6127* 	& 704.8(2) 	& 0.4  	& J1702$-$4305* 	& 538.4(5) 	& 0.2\\
J1123$-$6259	 & 223.4(1)	 & 0.1	& J1702$-$4310* 	& 377.6(3) 	& 0.2 \\
J1138$-$6207* 	& 520.4(4) 	& 0.3 	& J1705$-$1906 	& 22.94(2) 	& 0.01 \\
J1156$-$5707* 	& 243.2(1) 	& 0.2 	& J1705$-$3950* 	& 207.25(1) 	& 0.05 \\
J1216$-$6223* 	& 790(1) 	& 0.9 	& J1709$-$4429* 	& 75.68(3)	 & 0.02 \\ 
J1224$-$6407* 	& 97.686(4)	 & 0.006	 & J1715$-$3903* 	& 314.0(6)	 & 0.4 \\
J1248$-$6344* 	& 433.0(6) 	& 0.4 	& J1718$-$3825* 	& 247.46(6) 	& 0.04 \\
J1301$-$6305* 	& 374(1) 	& 1.00	 & 	J1722$-$3712*	& 99.49(3) 	& 0.01 	\\
J1302$-$6350* 	& 146.73(1) 	& 0.02  	& J1723$-$3659* 	& 254.4(1) 	& 0.07\\
J1305$-$6203* 	& 471.0(1) 	& 0.5	& J1726$-$3530* 	& 718(4) 	& 2  \\
J1320$-$5359	* & 97.1(1) 	& 0.08  	& J1730$-$3350* 	& 261.29(4) 	& 0.06  \\
J1327$-$6400	* & 679(1)	 & 0.7 	&  J1731$-$4744*	 & 123.056(4) 	& 0.005\\
J1341$-$6220* 	& 719.65(5) 	& 0.07 	& J1733$-$3716* 	& 153.18(8) 	& 0.05  \\
J1349$-$6130*	 & 284.5(1) 	& 0.1	& J1735$-$3258* 	& 758(3) 	& 1 \\
J1357$-$6429 	& 126(1) 	& 0.5	 &  J1737$-$3137* 	& 488.1(4) 	& 0.3\\
J1359$-$6038*	 & 293.736(3) 	& 0.004  	& J1738$-$2955 	& 222.5(6) 	& 1  \\
J1406$-$6121* 	& 537.8(4) & 0.5 	& J1739$-$2903* 	& 138.55(2) 	& 0.03 \\
J1410$-$6132* 	& 961.0(3) 	& 0.548	& J1739$-$3023* 	& 170.5(1) 	& 0.1 \\
J1412$-$6145* 	& 514.4(4) & 0.4 	 & J1740$-$3015* 	& 151.96(1) 	& 0.006 \\
J1413$-$6141* 	& 670.6(4) 	& 0.4	& J1745$-$3040 	& 88.112(7) 	& 0.007  \\
J1420$-$6048* 	& 360.15(6) 	& 0.1 	& J1756$-$2225* 	& 329(1) 	& 7  \\
J1452$-$5851*	 & 260.5(2) 	& 0.3 	&  J1757$-$2421	 & 179.38(2) 	& 0.02 \\
J1452$-$6036* 	& 349.54(2) 	& 0.02	 &  J1801$-$2154* 	& 386(1) 	& 0.5\\
J1453$-$6413*	 & 71.248(2) 	& 0.004	 & J1801$-$2304 	& 1070(1) 	& 0.7 \\
J1456$-$6843* 	& 8.639(7) 	& 0.008	 & J1801$-$2451* 	& 291.55(5)	 & 0.07 \\
J1509$-$5850*	 & 142.1(1) 	& 0.2 	&  J1803$-$2137	 & 234.01(5) 	& 0.03 \\
J1512$-$5759*	 & 627.47(1) 	& 0.02  	& J1806$-$2125* 	& 747(1)	 & 1 \\
\hline
\end{tabular}\end{table*}
\begin{table*}
\begin{tabular}{l c c | l c c}
\hline
Name & DM & dDM/d$t_{\rm limit}$ & PSR & DM & dDM/d$t_{\rm limit}$ \\
\hline
J1815$-$1738* 	& 724.6(2)	 & 0.3 	&  J1838$-$0549*	 & 276.6(4)	 & 0.5 \\
J1820$-$1529*	 & 768.5(6) 	& 0.7 	& J1839$-$0321* 	& 452.6(3) 	& 0.6\\
J1825$-$0935 	& 18.9(2) 	& 0.1	&   J1839$-$0905* 	& 344.5(3) 	& 0.7 \\
J1825$-$1446* 	& 352.23(4)	 & 0.05 	& J1841$-$0425 	& 325.13(3) 	& 0.02\\
J1828$-$1057* 	& 249(2) 	& 0.8	 & J1841$-$0524* 	& 284.5(3) 	& 0.5\\ 
J1828$-$1101* 	& 605.0(1) 	& 0.1	 & J1842$-$0905* 	& 343.4(2)	 & 0.1 \\
J1830$-$1059* 	& 159.70(1) 	& 0.01 	&  J1843$-$0355* 	& 797.7(6) 	& 0.7 \\
J1832$-$0827 	& 300.84(2) 	& 0.02 	&  J1843$-$0702* 	& 228.4(2) 	& 0.2   \\
J1834$-$0731* 	& 294.0(9)	 & 0.8	& J1844$-$0256* 	& 826(2) 	& 3 \\
J1835$-$0643* 	& 467.9(4) 	& 0.5 	&  J1844$-$0538* 	& 411.71(4) 	& 0.05 \\
J1835$-$0944* 	& 276.2(1) 	& 0.3 	& J1845$-$0434* 	& 230.8(2) 	& 0.1 \\ 
J1835$-$1106 	& 132.84(2) 	& 0.04 	& J1845$-$0743* 	& 280.93(2) 	& 0.01 \\
J1837$-$0559* 	& 319.5(6) 	& 0.5	& J1847$-$0402 	& 140.8(1) 	& 0.07 \\
J1837$-$0604* 	& 459.3(3)	 & 0.4  	&  J1853$-$0004* 	& 437.5(1) 	& 0.03  \\
J1838$-$0453*	 & 617.2(4)	 & 0.4	 & J1853+0011 	& 566(2) 	& 2  \\
\hline
\end{tabular}
\end{table*}

We note that a number of upper limits lie {\it below} the theoretical value expected from the median two dimensional velocity of 
338~kms$^{-1}$ \cite{hll+05}. This result is in contrast to the results
for the millisecond pulsars for which dDM/d$t$ is higher than expected.
By using the probability distribution
function for pulsar velocities given by \citet{hll+05} we estimate
that we should have had 12 pulsars in our sample with measurable values of dDM/d$t$, in good agreement with the 11 measured.

However, this velocity distribution may not be entirely accurate because it models pulsar's motion in 2-dimensional space. Detectable DM variations in some pulsars may be more realistically caused by irregularities along a single spatial axis in the interstellar turbulence \cite{bmg+10}. If this is the case we only need concern ourselves with a one dimensional velocity as the pulsar moves through the ISM. The probability of detecting pulsars in our sample can be recalculated using a one dimensional Gaussian distribution. We find a much lower detection estimate of 6 pulsars, including our real detections. While this alternate model of ISM turbulence may be the cause of variations along some of our lines of sight, we expect this effect only at low DM as the effects of multiple irregularities would cancel out over large distances.

We also note that our results appear to be at odds with \citet{bcc+04} who conclude that NE2001 underpredicts scattering, particularly at high DM. However, \citet{bcc+04} compare pulse-broadening times estimated from the \citet{cl02} model and their own CLEAN-based deconvolution algorithm, which are directly sensitive to an inner scale. Measurements of DM variations are not sensitive to the ISM inner scale which may explain the difference between the findings of these studies.

We find the two dimensional scattering modelled in NE2001 to agree well with the findings of our study, even in the high DM regime where that model becomes discrepant with others such as \citet{bcc+04}.

\section{Conclusions}\label{sec:conclusions}
We have analysed over five years of timing data for more than 160 young pulsars to search for any characteristic DM variations along several lines of sight. 

Only four pulsars in our sample PSRs J0835$-$4510, J0908$-$4913, J1824$-$1945, and J1833$-$0827 showed  highly significant changes in DM over the span of the study with seven other pulsars identified as marginal detections. One pulsar, PSR J1824$-$1945, displayed detectable DM variations at levels predicted by an interstellar medium dominated by Kolomogorov turbulence with no contribution from dense filaments local to the pulsar. DM variations for the Vela pulsar, PSR J0835$-$4510, were also consistent with a purely turbulent ISM, a dramatic change from measurements of large variations made 15 years ago attributed to the local SNR, indicating that perhaps the responsible filament is no longer moving through our line of sight. The other two detections, with variations well above those predicted by theory, are known to lie within turbulent local environments of supernova remnants or pulsar wind nebulae which make large contributions to observable turbulence along these particular lines of sight.

No DM variations were observed along the lines of sight to most pulsars in our sample, but we were able to set upper limits on detectable variations. We compared these limits with DM variations predicted from models of Kolmogorov turbulence and found our limits to be within an order of magnitude of theoretical predictions. Comparisons with accepted two dimensional velocity distributions using NE2001 scattering models suggested our experiment should have detected DM variations to approximately 12 pulsars. Confining our models to one dimension to simulate scattering effects due to irregularities in the interstellar turbulence reduces this estimate to 6 pulsars. We find our results to be in good agreement with 2D scattering from the NE2001 model.


The DM variations are a red process with a steep spectral exponent, therefore longer time baselines dramatically increase our sensitivity to DM variations. With more time the observation of young, high-DM pulsars will provide us with an excellent complement to the results obtained from millisecond pulsars at low DMs.

\section*{Acknowledgements} 

The Parkes radio telescope is part of the Australia Telescope which is funded by the Commonwealth of Australia for operation as a National Facility managed by CSIRO.\*
We thank the referee for the comments that improved the clarity of this manuscript.

\bibliographystyle{mnras}
\bibliography{dmrefs,journals}

\end{document}